\documentclass[english,aps,amssymb,prl,twocolumn,showpacs]{revtex4-1}
\pdfoutput=1
\usepackage[T1]{fontenc}
\usepackage[latin9]{inputenc}
\usepackage{amsmath}
\usepackage{graphicx}
\usepackage{amssymb}
\usepackage{esint}

\makeatletter
\@ifundefined{textcolor}{}
{
 \definecolor{WHITE}{gray}{1}
 \definecolor{RED}{rgb}{1,0,0}
 \definecolor{GREEN}{rgb}{0,1,0}
 \definecolor{BLUE}{rgb}{0,0,1}
 \definecolor{CYAN}{cmyk}{1,0,0,0}
 \definecolor{MAGENTA}{cmyk}{0,1,0,0}
 \definecolor{YELLOW}{cmyk}{0,0,1,0}
 }

\renewcommand{\phi}{\varphi}
\renewcommand{\epsilon}{\varepsilon}

\renewcommand{\vec}[1]{{\bf #1}}
\newcommand{Û}{$}

\usepackage{epsfig}
\@ifundefined{showcaptionsetup}{}{%
\PassOptionsToPackage{caption=false}{subfig}}
\usepackage{subfig}
\makeatother

\usepackage{babel}

\begin{document}

\title {Topological pi Josephson junction in superconducting Rashba wires}

\author{Teemu Ojanen}
\email[Correspondence to ]{teemuo@boojum.hut.fi}
\affiliation{Low Temperature Laboratory (OVLL), Aalto University, P.~O.~Box 15100,
FI-00076 AALTO, Finland }
\date{\today}
\begin{abstract}
In this work we show that Rashba-based topological superconductor nanowires, where the spin-orbit coupling may change its sign, support three topological phases protected by chiral symmetry. When a superconducting phase gradient is applied over the interface of the two nontrivial phases, the Andreev spectrum is qualitatively phase shifted by $\pi$ compared to usual Majorana weak links.  The topological $\pi$-junction has the striking property of exhibiting maximum supercurrent in the vicinity of vanishing phase difference. Qualitative features of the junction are robust against disorder and magnetic fields violating chiral symmetry. The studied system could be realized by local gating of the wire or by an appropriate stacking of permanent magnets in synthetic Rashba systems.

\end{abstract}
\pacs{73.63.Nm,74.50.+r,74.78.Na,74.78.Fk}
\maketitle
\bigskip{}

\emph{Introduction}-- Recent experimental evidence is consistent with the possibility that superconducting nanowires with a strong spin-orbit coupling support topological superconductivity and Majorana bound states (MBS) \cite{mourik, das, deng}. Though the existence of MBS is not conclusively settled yet, experimental efforts have increased the interest towards the topic significantly. Besides being interesting due to their exotic physical properties, MBS could find applications in quantum information processing. \cite{nayak, kitaev1, kitaev2, kitaev3, alicea2}.

Theoretical developments in topological insulator physics \cite{fu, kane2} and two-dimensional semiconductor heterostructures \cite{sato, lutchyn1} paved the way for the proposal of  topological superconductivity in nanowires with a strong Rashba spin-orbit coupling \cite{lutchyn, oreg}.  The topological phases of nanowire systems are commonly classified assuming only particle-hole symmetry leading to a $Z_2$ invariant. However, in the special case when the applied magnetic field is perpendicular to the effective spin-orbit field and the superconducting order parameter is uniform, the system possesses additional chiral symmetry. Then the system may have an arbitrary number of distinct topological phases characterized by a $Z$-valued invariant \cite{volovik, schnyder, tewari}. Since unbroken chiral symmetry relies on fine tuning of the magnetic field direction, classification of the nanowire system is usually regarded as $Z_2$ type.

In this work we show that the concept of chiral symmetry in nanowires is not merely a mathematical curiosity and can have distinct physical consequences. We consider a nanowire system where in some segments of the wire the Rashba coupling constant may have opposite signs and show that this system possesses three topological phases protected by chiral symmetry. The three phases are distinguished by an invariant taking values $\nu=\pm 1$ and $\nu=0$. The phases $\nu=\pm 1$ correspond to the nontrivial phase in the $Z_2$ classification while $\nu=0$ phase coincides with the trivial phase. The interface between $\pm 1$ and $0$ is similar to the boundary between trivial and nontrivial phases in $Z_2$ classification but a boundary between $1$ and $-1$ phases contains two degenerate MBS protected by chiral symmetry. As discussed below, the sign change of the Rashba coupling could be realized, for example, by local gating of the wire. Alternatively,  in synthetic Rashba systems based on spatially varying magnetization, an appropriate stacking of magnets in the proximity of the wire could implement the sign reversal.

When a superconducting phase difference is applied over the phase boundary where the Rashba coupling changes sign, chiral symmetry is violated and the degeneracy of the two MBS in the junction is lifted. The two MBS are degenerate when the phase difference is $2n\pi$, in contrast to usual Majorana weak links, spectrum of which has a pair of degenerate MBS when the phase difference is $(2n+1)\pi$ with an integer $n$. Thus, the sign change of the Rashba coupling qualitatively amounts to a  phase shift of $\pi$ of the Andreev spectrum. Due to special properties of a topological Josephson junction, the spectrum of a topological $\pi$-junction supports maximum supercurrent around vanishing phase difference.

We begin by introducing the nanowire model and classifying phases according to chiral symmetry. The MBS wavefunctions on the $-1/1$ phase boundary are solved analytically in special cases. The Andreev spectrum of a topological $\pi$-junction is studied numerically by solving a tight-binding model of a finite nanowire, including impurity potential and magnetic fields violating chiral symmetry. Qualitative features of the junction are insensitive to the detailed manner how the Rashba coupling changes sign.
\begin{figure}[t]
\centering
\includegraphics[width=0.40\columnwidth, clip=true]{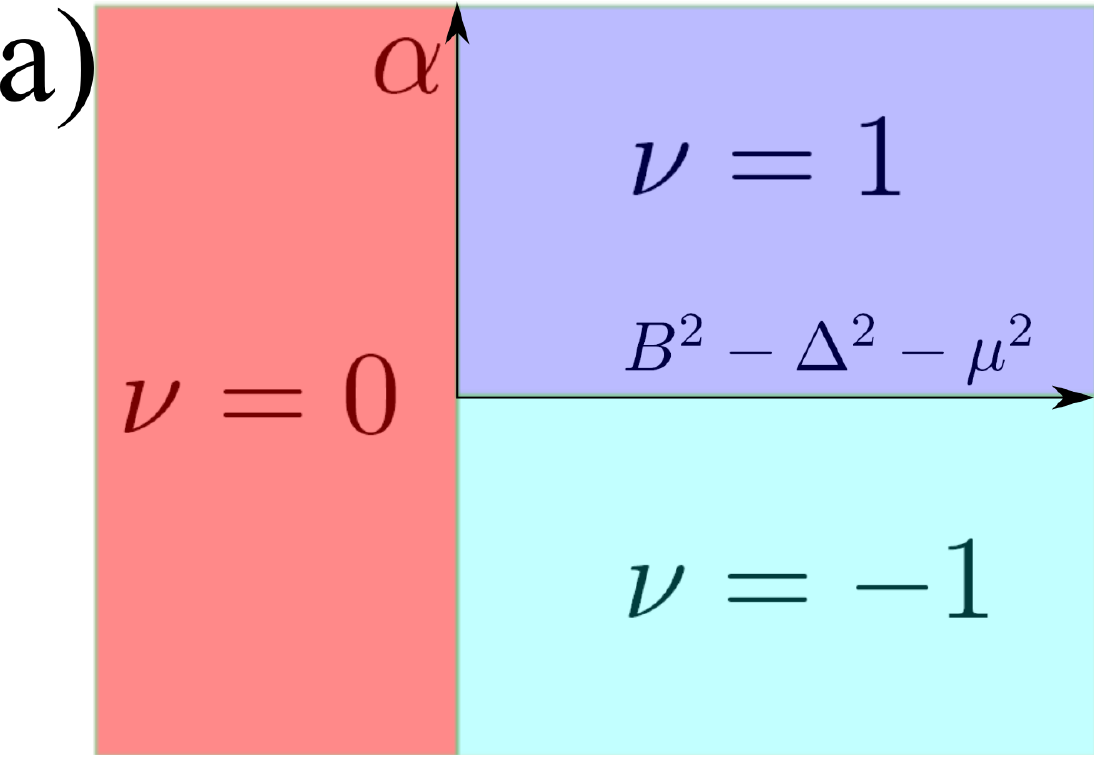}
\includegraphics[width=0.50\columnwidth, clip=true]{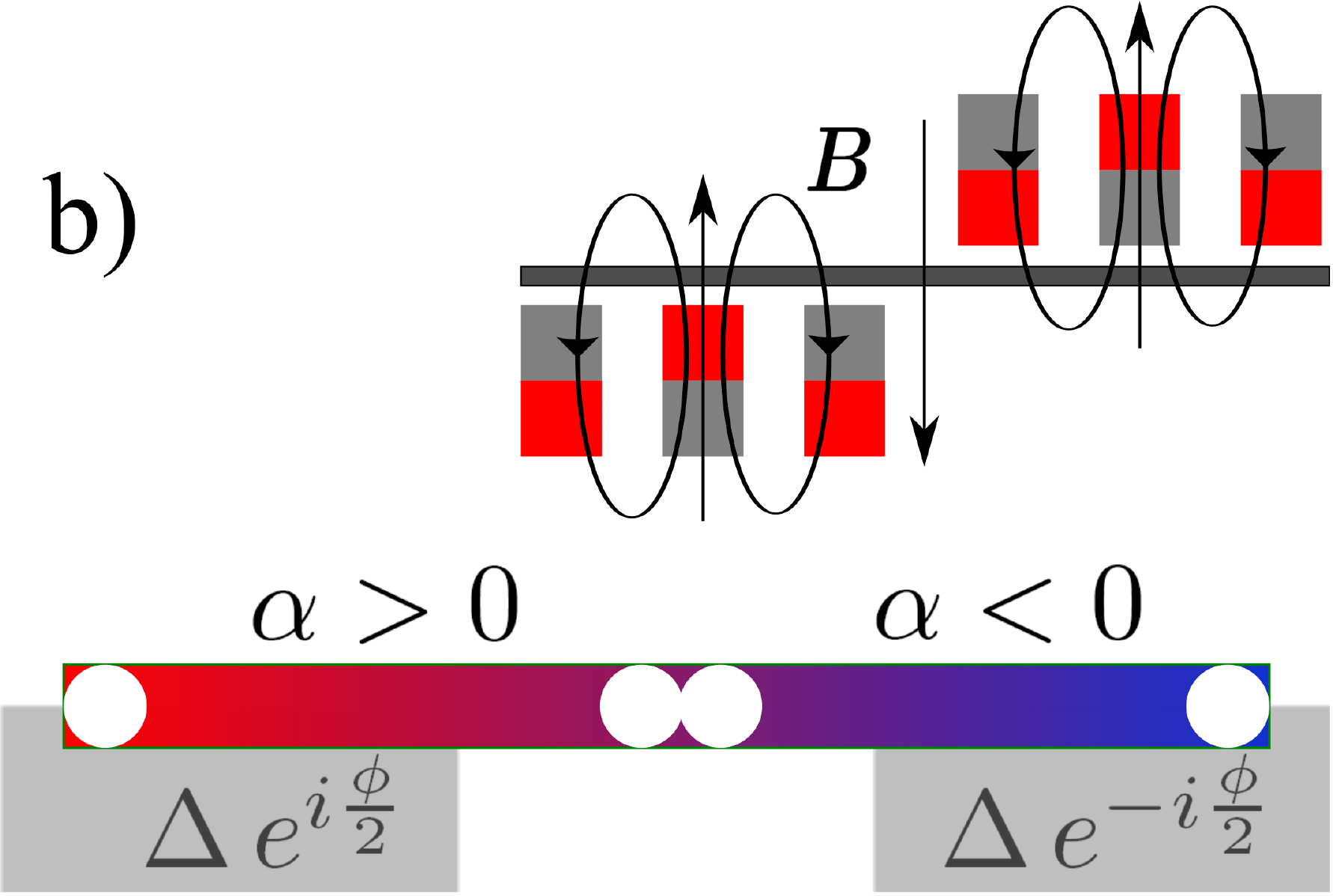}
\caption{a):  Phase diagram of the nanowire system in Û(B^2-\Delta^2-\mu^2,\alpha)Û-plane in the presence of chiral symmetry. b): Schematic representation of the topological Û\piÛ-junction formed between wire segments where the Rashba coupling changes sign. The interface contains two Majorana bound states that become (nearly) degenerate  at the vanishing superconducting phase difference.  Inset: Magnetic realisation of an effective sign-changing Rashba coupling.    }
\label{sceme}
\end{figure}

\emph{Model and phase diagram}--
We study a wire with a Rashba spin-orbit coupling in the vicinity of a superconductor in magnetic fields.
The system is modeled by the Bogoliubov-de Gennes Hamiltonian \cite{lutchyn, oreg},
\begin{align}\label{h}
H(k)=&\left(\epsilon_k+\alpha k\sigma_x\right)\tau_z+\boldsymbol{B}\cdot\vec{\sigma} +\vec{\Delta}\cdot\vec{\tau},
\end{align}
where $\epsilon_k=\frac{\hbar^2k^2}{2m}-\mu$ and $\sigma_i$ and $\tau_i$ are Pauli matrices operating in the spin and the Nambu space, respectively. The Hamiltonian (\ref{h}) is expressed in the basis  $\Psi=(\psi_{\uparrow}, \psi_{\downarrow},\psi^{\dagger}_{\downarrow},-\psi^{\dagger}_{\uparrow} )^T$ and the coordinates have been chosen so that the wire is parallel to the $y$ direction and the effective spin-orbit field coincides with the $x$ direction. The first term represents kinetic energy of electrons and holes including the Rashba coupling characterized by a coupling constant $\alpha$. The second term is the Zeeman coupling $\vec{B}=(B_x,B_y,B_z)$ due to the applied magnetic field and the last term arises from the proximity-induced superconducting pairing with $\vec{\Delta}=\Delta(\mathrm{cos}(\phi)\tau_x,\mathrm{sin}(\phi)\tau_y,0)$.

In the special case when the magnetic field is perpendicular to the spin-orbit field and the superconducting phase is uniform in the system, one can choose $\vec{B}=(0,0,B_z)$ and $\vec{\Delta}=(\Delta,0,0)$. Then the Hamiltonian anticommutes $\{H,C\}=0$ with  $C=\sigma_{x}\tau_{y}$, thus possessing chiral symmetry. As a consequence, if $H$ has an eigenstate $|E_{i}^{+}\rangle$ with energy $E_{i}$, then there exists another eigenstate $C|E_{i}^{+}\rangle$ belonging to energy $-E_{i}$. Zero modes, when present, can be chosen to coincide with the eigenstates of the generator of chiral symmetry, $C|\psi^{\pm}\rangle=\pm|\psi^{\pm}\rangle$. In odd spatial dimensions systems with chiral symmetry posses
topological phases characterized by a $Z$-valued topological invariant \cite{schnyder, volovik, tewari}. There exists different equivalent formulations of the invariant, here we consider a form appropriate for 1d systems
\begin{align}
\nu[H]=\frac{1}{4\pi i}\int dk\,\mathrm{Tr}\left[CH^{-1}\partial_{k}H\right], \label{wind}
\end{align}
which was recently employed in Ref.~\cite{vayrynen} to analyse similar insulating systems. The phase diagram of the system in the $( B^2-\Delta^2-\mu^2, \alpha)Û plane is depicted in Fig.~1 a), revealing three distinct phases $\nu=\pm 1$ and $\nu=0$.  The $\nu=0$ phase, corresponding to the parameter regime $B^2<\Delta^2+\mu^2$, can be identified with the trivial phase in the $Z_2$ classification. The $\nu=\pm 1$ phases differ by the sign of the Rashba coupling and map to the nontrivial phase in the $Z_2$ scheme which is insensitive to the sign of $\alpha$. Thus the phase boundaries between $\nu=\pm 1$ and $\nu=0$ phases behave exactly as the $Z_2$ interface between the trivial and the nontrivial phase supporting a single MBS \cite{oreg}. However, the boundary between $\nu=-1$ and $\nu=1$ is expected to be qualitatively different and, in fact, it supports two MBS. These MBS are degenerate at zero energy and cannot be fused to form finite energy states as long as chiral symmetry remains intact.

The existence of two MBS can be demonstrated by direct analytical solution of the zero modes on the $-1/1$-phase boundary. To simplify the analysis, we assume that the Rashba coupling $\alpha(x)$ is negative constant for $x<0$ and positive constant for $x>0$ and that the parameters satisfy $\mu,\Delta<B\ll E_{R}\equiv m\alpha^2/2\hbar$. Then we can linearize (\ref{h}) in $k$ and solve the zero mode wave functions from   $\left[-i\alpha(x)\sigma_x\tau_z\partial_x-\mu\tau_z+B\sigma_z+\Delta\tau_x\right]\psi=0$ \cite{oreg}. Matching the solutions for different signs of $\alpha$ at $x=0$ we recover two normalizable solutions
\begin{align}\label{mbs}
\psi_{1/2}&=A_{1/2}e^{\frac{x}{\alpha(x)}(-\sqrt{B^2-\mu^2}\pm\Delta^2 )}\nonumber\\
&\frac{e^{\pm i\frac{\pi}{4}}}{2}\left[e^{i\frac{\theta}{2}}|x_+\rangle|y_\pm\rangle+e^{-i\frac{\theta}{2}}|x_-\rangle|y_\mp\rangle\right],
\end{align}
where $e^{i\theta}=\frac{\mu}{B}-i\frac{\sqrt{B^2-\mu^2}}{B}$ and $A_{1/2}$ are real constants (here $B^2>\Delta^2+\mu^2$). The vectors $|x_\pm\rangle$ and $|y_\pm\rangle$ correspond to the eigenstates of operators $\sigma_x$ and $\tau_y$ and the upper (lower) signs on the right-hand side of Eq.~(\ref{mbs}) correspond to $\psi_1$ ($\psi_2)$. The zero modes satisfy $C\psi_{1/2}=\pm\psi_{1/2}$ and can be obtained analogously for the opposite boundary where $\nu=1$ changes to $\nu=-1$. Expressing the wavefunctions in a standard four component form, the corresponding field operators can be written as $\gamma_{1/2}=\psi_{1/2}\cdot\Psi$, which satisfy $\gamma_{1/2}=\gamma_{1/2}^\dagger$. Although the analytical MBS solutions were obtained in a special parameter regime, the number of solutions cannot change as long as the bulk gaps are finite and chiral symmetry remains unbroken.

\emph{Topological $\pi$-junction: general discussion}-- Different aspects of the Josephson effect in topological superconductors are under active research \cite{chung, sanjose, pikulin, jiang1, jiang2, meng, domingues, kotetes}. The $4\pi$ periodic Josephson effect between two topological p-wave segments coupled through a weak link was already discussed in the pioneering work by Kitaev \cite{kitaev1}. Typically the Andreev spectrum of a  topological nanowire weak link contains two MBS states that become degenerate  when the superconducting phase difference Û\phi=(2n+1)\piÛ is applied over the junction. This is qualitatively in striking contrast to the above discussed behaviour  the $\nu=\pm1$ interface which supports degenerate MBS at Û\phi=n2\piÛ. This suggests that spectrum of the  $\nu=\pm1$ junction is qualitatively phase shifted by Û\piÛ. 

As discussed by a number of authors, Hamiltonian (\ref{h}) can be expressed in the basis $\psi_{\pm}$ of eigenfunctions of the system in the absence of superconductivity (see for example Refs. \cite{alicea1,potter}). At low filling only the lower band $\psi_{-}$ is populated and the effective gap function projected to this band is given by $\Delta_p=\frac{\alpha k}{\sqrt{\alpha^2k^2+B^2}}\Delta$.  This expression suggests that changing the sign of the Rashba coupling $\alpha$ is effectively equivalent to imposing a rigid phase shift of $\pi$ to the order parameter. Therefore we expect that the phase boundary between the $\nu=1$ and $\nu=-1$ phases effectively behaves as a $\pi$-junction when a finite phase difference is applied. The topological $\pi$-junction has an interesting difference compared to the usual $\pi$-junctions. In the usual case the $\pi$ phase shift replaces a local minimum of the original spectrum at $\phi=0$ (mod Û2\piÛ) by a local maximum. Since the current through a short junction is  given by $I(\phi)=\frac{2e}{\hbar}\partial_\phi E(\phi)$, where ÛE(\phi)Û is the energy of the populated Andreev level,  supercurrent  vanishes at $\phi=0$ for a topologically trivial Û\piÛ-junction. In the topological $\pi$-junction the $\pi$ phase shift moves the zero-energy crossing from $\phi=\pi$ to $\phi=0$ (mod Û2\piÛ). The vicinity of the crossing point is not an extremum of the spectrum, therefore giving rise to a spontaneous supercurrent $I(0)\neq 0$ in infinite systems.  In a finite system the picture is slightly modified since the crossing at Û\phi=0Û becomes avoided due to coupling to the MBS at the end of the wire (depicted in Fig. 1 b) ) and supercurrent vanishes $I(0)=0$. Avoided crossing has the same origin as in usual topological Josephson junctions \cite{sanjose, pikulin}. However, the avoided crossing becomes exponentially narrow when the wire is longer than the localisation length of MBS wavefunctions, thus yielding maximum supercurrent in the immediate vicinity of Û\phi=0Û.   


\emph{Tight-binding calculation of $\pi$-junction spectrum}--
Above we gave general arguments why we expect $\nu=\pm 1$ interface to behave as a Û\piÛ-junction.
Here we study specific properties of the topological Û\piÛ-junction by numerically solving the spectrum of a tight-binding model of the setup in Fig.~\ref{sceme} b). The Hamiltonian of a lattice model is given  by
\begin{align} \label{tb}
H=&-\sum_{\langle i, j \rangle} ( t\,\Psi_i^{\dagger}\tau_z\Psi_j+\mathrm{h.c})+\sum_{i }(-\tilde{\mu}  \Psi_i^{\dagger}\tau_z\Psi_i+ V_i \Psi_i^{\dagger}\tau_z\Psi_i)\nonumber\\
&+\sum_{i }  \Psi_i^{\dagger}\vec{B}\cdot\vec{\sigma}\Psi_i+ \frac{i}{2a}\sum_{\langle i, j \rangle}\alpha_i ( \Psi_i^{\dagger}\sigma_x\tau_z\Psi_j-\mathrm{h.c})\nonumber\\
&+\sum_{i }  \Delta_i(\mathrm{cos}\,\phi_i\Psi_i^{\dagger}\tau_x\Psi_i+\mathrm{sin}\,\phi_i\Psi_i^{\dagger}\tau_y\Psi_i),
\end{align}
where Û\Psi_iÛ is a four-component spinor at lattice site ÛiÛ and ÛaÛ is the lattice constant. The terms proportional to ÛtÛ and Û\tilde{\mu}Û on the right-hand side of (\ref{tb}) implement the term proportional to Û\epsilon_kÛ in Eq.~(\ref{h}), while the terms on the second line accommodate the Zeeman splitting and the Rashba coupling. The terms on the third line corresponds to the superconducting order parameter which  is spatially varying. The normal and Rashba hopping terms couple only nearest-neighbour sites. We have allowed a spatially varying Rashba coupling and included a random potential ÛV_iÛ to study the effects of disorder. In the following we are working in the units where energies are expressed in the units of the Rashba scale ÛE_R=m\alpha^2/2\hbar^2Û and the lengths are given in the units of the spin-orbit length Ûl_{\mathrm{so}}=2\hbar^2/m\alphaÛ (here $\alpha$ corresponds to the absolute value of the bulk Rashba coupling away from the junction).  Typical values for these parameters InSb wires are ÛE_R\sim 50Û Û\mu eVÛ and Ûl_{\mathrm{so}}\sim200Û nm \cite{mourik}.  The disorder potential is modelled as a (discrete) Gaussian white noise with a correlation function Û\langle V_iV_j \rangle=\frac{\hbar^2v_F^2}{l_ea}\delta_{ij}  Û  characterized by the elastic mean free path Ûl_eÛ and the Fermi velocity  $v_F=E_{R}l_{\mathrm{so}}/\hbar$. We are interested in the energy spectrum of a superconducting-normal wire-superconducting  (SNS) junction where the left S segment is in the Û\nu=-1Û phase and the right one is in the Û\nu=1Û phase.  Other relevant length scales in the problem are the superconducting coherence length Û\xi=\hbar v_F/\DeltaÛ, the length of the normal segment ÛL_NÛ and  the length scale Û\xi_RÛ associated with the  sign change of the Rashba coupling. In the following we model the variable Rashba coupling by Û\alpha_j=\alpha\,\mathrm{tanh}\,\left[(j-N/2)a/\xi_R\right]+\alpha'Û with Û|\alpha|>|\alpha'|Û.  Since we are working with a finite system of ÛNÛ lattice sites, the energy level crossings of topological origin in an infinite system are rendered to weakly avoided crossings that approach degeneracy as the effective system size is increased. Motivated by  experiments  \cite{mourik}, we consider wires with effective lenghts ÛL\gtrsim 10\,l_{\mathrm{so}}Û (Û\simÛ 2-4 Û\muÛm). 

\begin{figure}[t]
\centering
\includegraphics[width=0.45\columnwidth, clip=true]{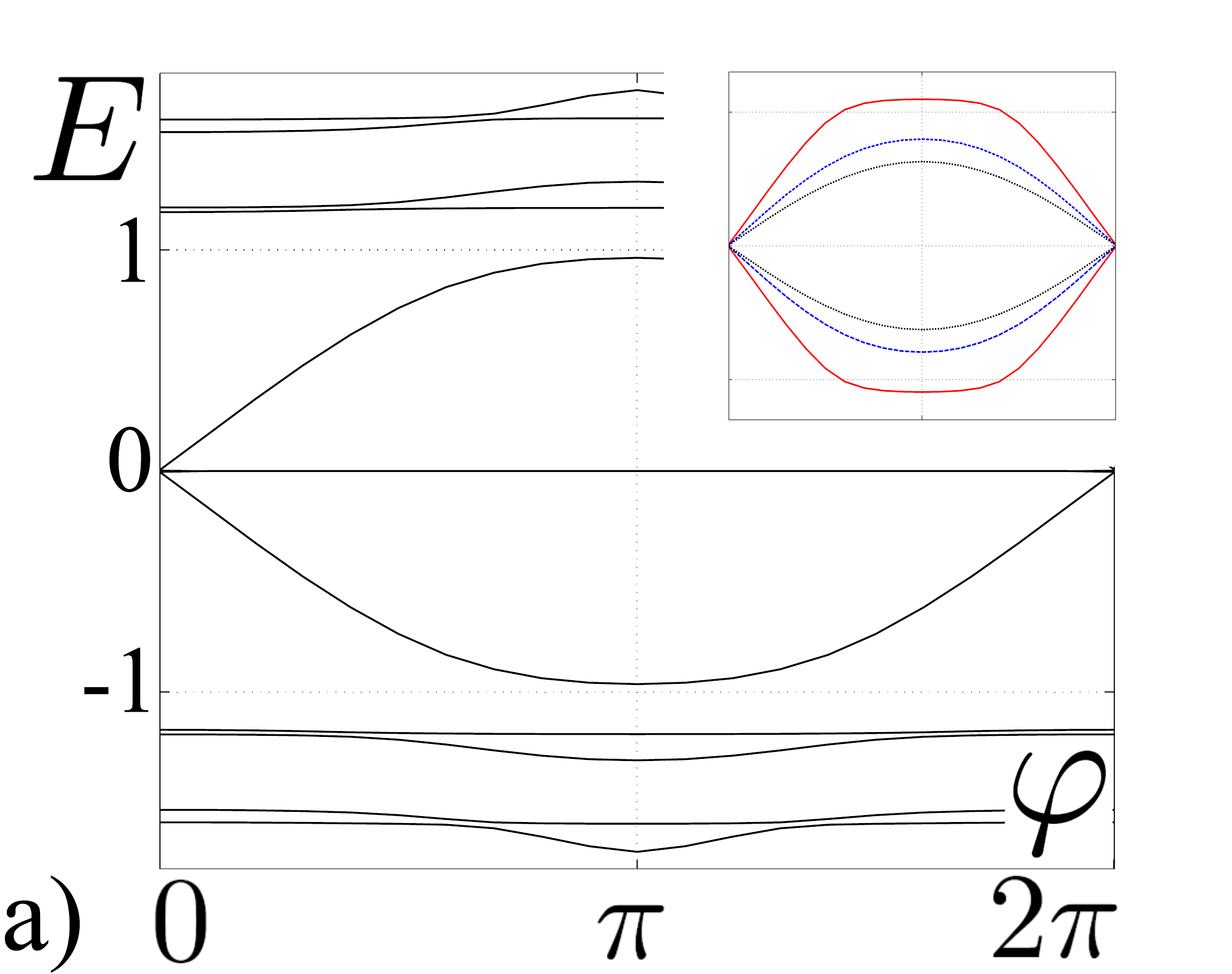}
\includegraphics[width=0.45\columnwidth, clip=true]{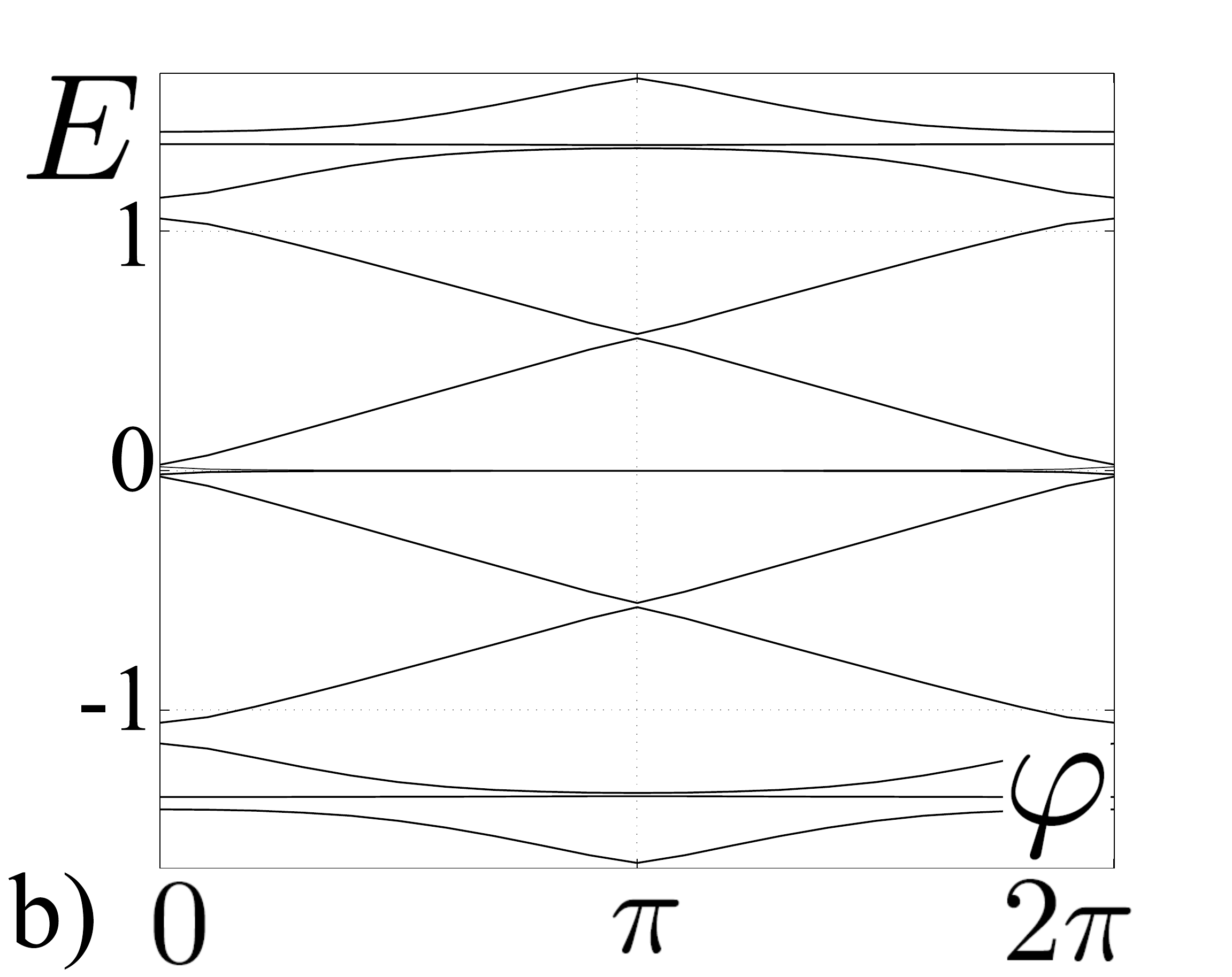}
\caption{a): Low-lying spectrum of a wire with a short junction (ÛL_N\ll \xiÛ) as a function of the phase difference, corresponding to parameters ÛB=4Û, Û\Delta=3Û, Û\mu=0Û and Ûl_e=\inftyÛ. The effective length of the wire is ÛL=10l_{\mathrm{so}}Û, Û\xi_R=0.5\,l_{\mathrm{so}}Û and ÛN=200Û. The flat mid-gap level is doubly degenerate, corresponding to the Majorana end states. The arcs correspond to the MBS at the junction, being nearly degenerate at Û\phi=0,2\piÛ. Inset: MBS spectrum at the junction for Û\xi_R=0.1\,l_{\mathrm{so}}Û, Ûl_{\mathrm{so}}Û and Û2\,l_{\mathrm{so}} Û (outside to inside).    b):  Same as  a) in the case of a long junction ÛL_N=4l_{\mathrm{so}}=12\xiÛ.  }
\label{juttuja}
\end{figure}

 Spectra of  wires in the clean limit ÛV_i=0Û in the case Û\vec{B}= (0,0,B)Û are shown in Fig.~\ref{juttuja}. At Û\phi=0,2\piÛ chiral symmetry is unbroken and the Andreev states corresponding to the MBS at the junction are degenerate within the resolution of the plot. The spectrum of a short junction $L_N\ll\xi$, depicted in Fig.~2 a), has two bound states localized at the junction and qualitatively corresponds to a  Û\piÛ-shifted spectrum of usual Majorana weak links \cite{kitaev1,oreg,lutchyn}. The inset illustrates how the spectrum of MBS at the junction depends on the width of the Rashba domain wall Û\xi_RÛ.  Longer junctions $L_N>\xi $ may have a number of  states localized on the junction, as shown in Fig.~\ref{juttuja} b), but also in that case spectrum is qualitatively shifted by Û\piÛ. Qualitative features of spectra are insensitive to the relative lengths of  $L_N$ and $\xi_R$. 

When magnetic field has a finite overlap with the spin-orbit field $B_x\neq 0$, chiral symmetry is violated and the strict topological distinction between $\nu=\pm1$ phases disappears. This also implies that the (near) degeneracy of the MBS at $\phi=0$ is lifted and moves to $\phi_0\neq0$. Remarkably, as numerical calculations indicate, the violation of chiral symmetry by finite $B_x$ is inefficient in lifting the degeneracy when the Rashba domain wall is not strongly asymmetric Û\alpha'\ll\alphaÛ. As shown in Fig.~3 a), the offset is small $\phi_0\approx 0$ even for large symmetry-breaking fields $B_x\sim E_R$  and significant asymmetry $(\alpha+\alpha')/(\alpha-\alpha')= 1.5$. Thus, the operation of the $\pi$-junction \emph{does not} require a special fine tuning of parameters. This behaviour is markedly different from the case of Majorana weak links where the sign of the Rashba coupling is constant. There the $B_x$-induced shift of the crossing point, taking place at odd multiples of $\pi$ when ÛB_x=0Û, is large $(\phi_0-\pi) \,\mathrm{mod}\, 2\pi\sim 1$ for corresponding values of $B_x$.

In Fig.~3 b) we have plotted $\pi$-junction spectra in the presence of disorder. For a realistic disorder strength $l_e=2\,l_{\mathrm{so}}$ the qualitative picture of clean wires survive although the spectra exhibit significant sample to sample fluctuations. Potential disorder affects the amplitude of the dispersion but cannot move the (avoided) crossing points. In the strong disorder regime $l_e\ll\,l_{\mathrm{so}}$ the well-defined MBS localized at the junction are absent  and the picture discussed in the clean limit breaks down.
\begin{figure}[t]
\centering
\includegraphics[width=0.45\columnwidth, clip=true]{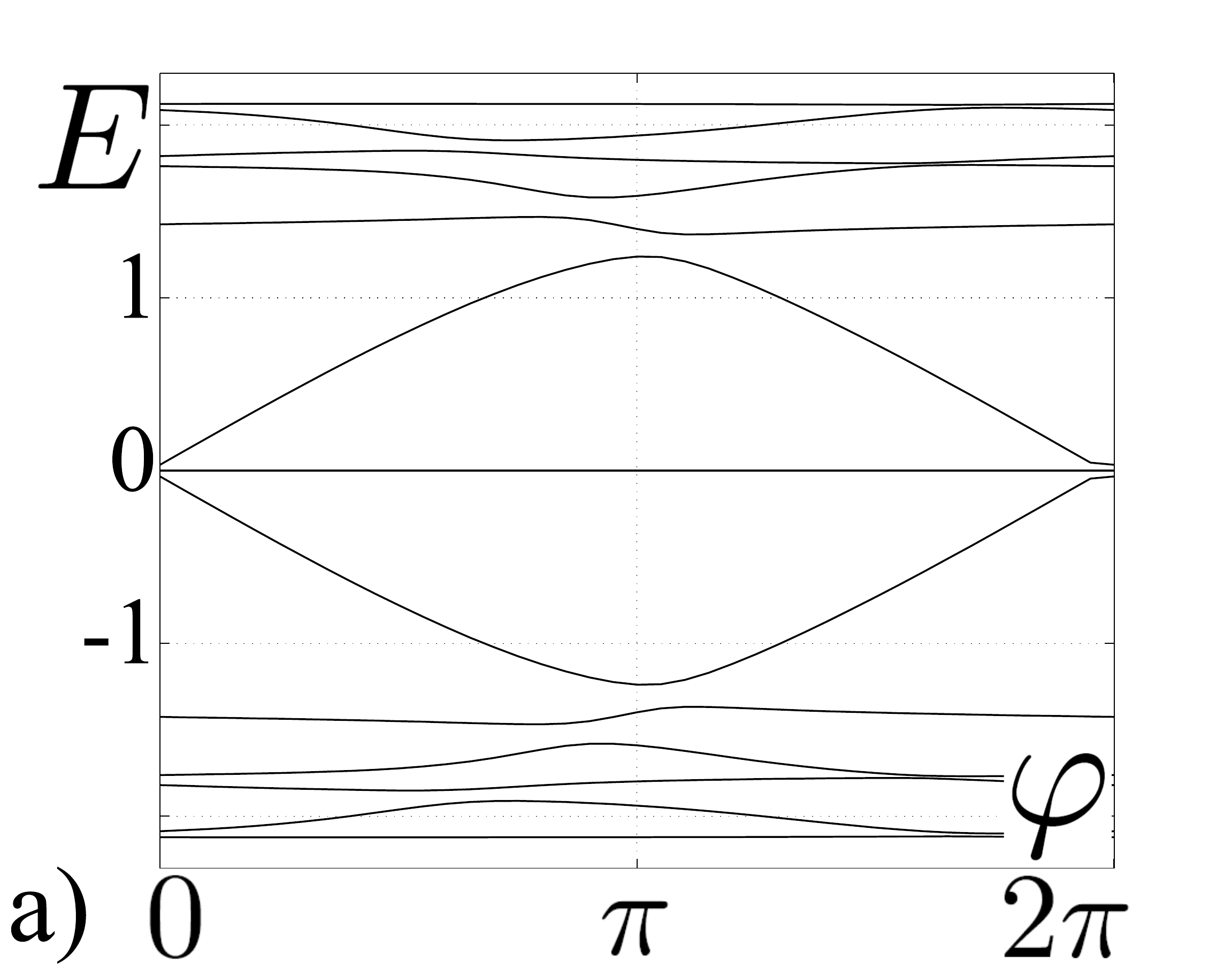}
\includegraphics[width=0.47\columnwidth, clip=true]{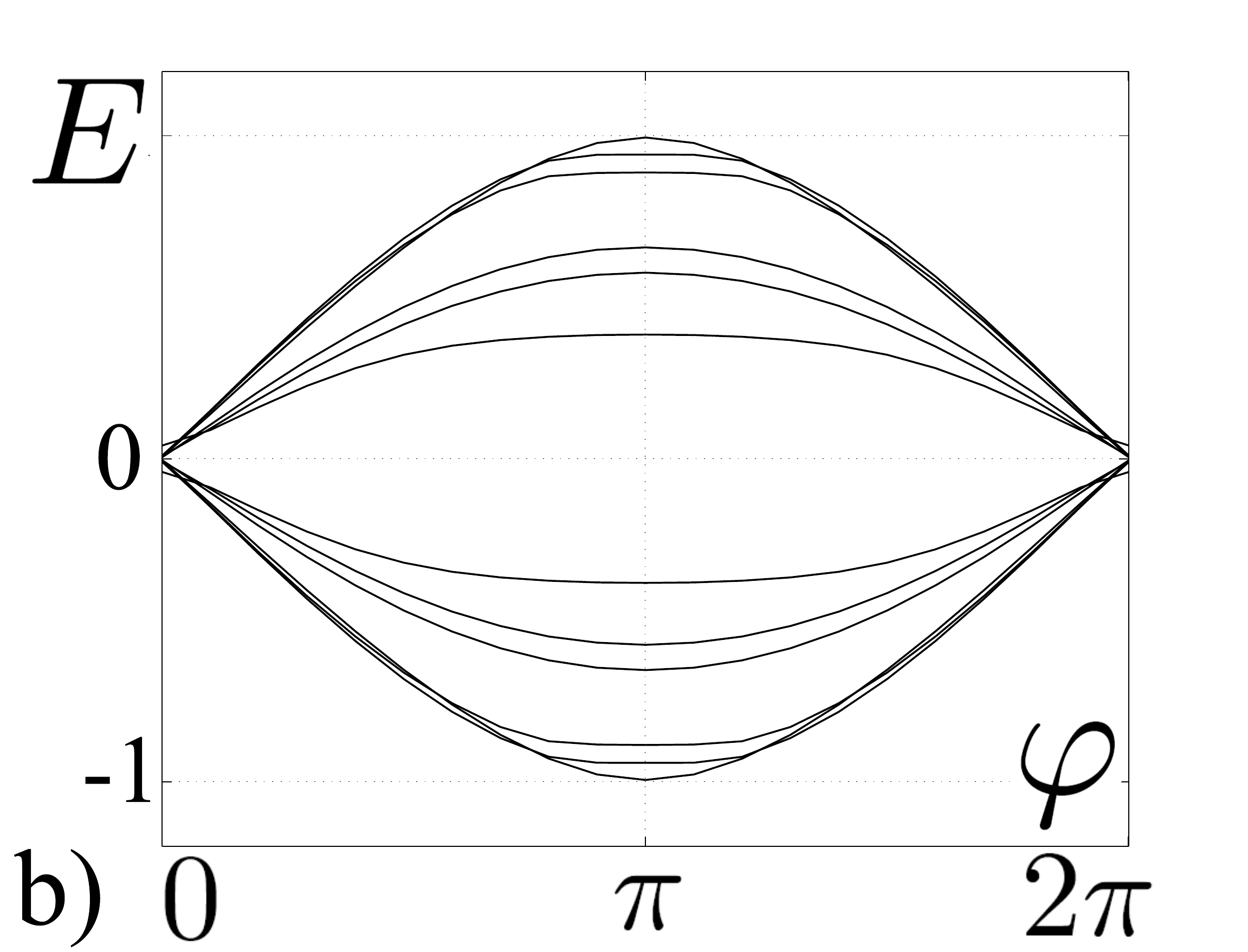}
\caption{a): Spectrum of a wire with a short junction and asymmetric Rashba domain wall Û\alpha'/\alpha=0.2Û in the presence of large chiral symmetry breaking field ÛB_x=1Û, other parameters  ÛB=5Û, Û\Delta=3Û, Û\mu=0Û and Ûl_e=\inftyÛ, ÛL=13l_{\mathrm{so}}Û,  Û\xi_R=0.5\,l_{\mathrm{so}}Û and ÛN=200Û.    b): MBS spectrum of a short junction in the presence of disorder. The different curves correspond to different disorder configurations with Ûl_e=2l_{\mathrm{so}}Û, other parameters  ÛB=5Û, Û\Delta=3Û, Û\mu=1Û, ÛL=10\,l_{\mathrm{so}}Û,  Û\xi_R=0.5\,l_{\mathrm{so}}Û and ÛN=200Û.      }
\label{jotain}
\end{figure}

\emph{Discussion--} In 2d semiconductor structures a finite Rashba coupling can be traced to structural asymmetry, resulting to a finite average electric field perpendicular to the sample plane. Moving electrons thus experience a magnetic field which couples to their spin. Structural asymmetry can be affected by asymmetric doping and applying a gate voltage. A gate-induced sign reversal has been demonstrated in 2d systems \cite{studer} which is encouraging for the present proposal. Semiconductor nanowire experiments show that the direction of the effective spin-orbit field is perpendicular to the wire and the substrate plain \cite{mourik, nadj}.  The sign of the Rashba coupling could be tuned, in principle, by local gating in wires also. Since the location of chemical potential needs to be controlled to achieve a topological segment, two independent gates are required. Although the double-gate tuning of the Rashba coupling in InAs wires has been experimentally realized  \cite{liang}, combining this method  with proximity superconductivity is a challenging task.

Recently it was proposed that the Rashba coupling in nanowires could be emulated by spatially varying magnetization \cite{braunecker1,braunecker2, kjaergaard}. As discussed in Ref.~\cite{kjaergaard}, an effective Rashba coupling could be realized by a realistic array of nanomagnets in the vicinity of the wire. The effective Rashba coupling is proportional to the magnetic field gradient along the wire and may take values comparable to couplings in intrinsic Rashba wires \cite{kjaergaard}. In a configuration where the field is restricted to the sample plane, the sign of the Rashba coupling is determined by the rotation direction of the field along the wire \cite{kjaergaard}.  If the magnets are stacked  as shown in the inset of Fig. \ref{sceme} b),  the rotation of the magnetic field along the wire changes from clockwise to counterclockwise in the middle, effectively implementing the sign change of the Rashba coupling. Synthetic Rashba couplings are typically spatially modulated which, according our numerical calculations, does not affect the operation of the $\pi$-junction significantly. Thus, engineering a topological $\pi$-junction by employing the method proposed in Ref.~\cite{kjaergaard} should not require higher sophistication than realizing a wire with a coupling with a uniform sign.

\emph{Conclusion--}
By controlling the sign of the Rashba coupling constant in superconducting nanowires, it is possible to realize three topological phases protected by chiral symmetry. The interface between the two distinct topologically nontrivial phases supports two degenerate Majorana bound states and behaves as a topological $\pi$-junction when a superconducting phase gradient is applied over the interface. This junction exhibits maximum supercurrent in the vicinity of vanishing phase difference and is robust against disorder and insensitive to moderate chiral symmetry breaking magnetic fields.

It is pleasure to thank Jay Sau and Pauli Virtanen for discussions. The author acknowledges Academy of Finland for support.

\end{document}